\documentclass[]{aa}

\usepackage{graphicx}
\usepackage{txfonts}
\usepackage{wasysym}
\usepackage{pgf}
\usepackage{booktabs}

\begin{document}

   \title{Interaction of infalling solid bodies with primordial atmospheres of disk-embedded planets}

   \author{Florian Ragossnig \inst{1}
      \and Alexander St{\"o}kl \inst{1}
      \and Ernst Dorfi \inst{1}
          \and Colin P. Johnstone \inst{1}      
      \and Daniel Steiner \inst{1}
      \and Manuel Güdel \inst{1}}

   \institute{Institute for Astrophysics, University of Vienna, T\"urkenschanzstrasse 17, A-1180 Vienna, Austria}

\authorrunning{F. Ragossnig et al.}
\titlerunning{TBD}

\offprints{F. Ragossnig}

\date{Received 22/01/2018; accepted 19/06/2018}

  \abstract
   {Planets that form early enough to be embedded in the circumstellar gas disk accumulate thick atmospheres of nebular gas. Models of these atmospheres need to specify the surface luminosity (i.e. energy loss rate) of the planet. This luminosity is usually associated with a continuous inflow of solid bodies, where the gravitational energy released from these bodies is the source of energy. However, if these bodies release energy in the atmosphere instead of at the surface, this assumption might not be justified.}
   {Our aim is to explore the interactions of infalling planetesimals with primordial atmospheres at an embedded phase of evolution. We investigate effects of atmospheric interaction on the planetesimals (mass loss) and the atmosphere (heating/cooling).}
   {We used atmospheric parameters from a snapshot of time-dependent evolution simulations for embedded atmospheres and simulated purely radial, infall events of siliceous planetesimals in a 1D, explicit code. We implemented energy transfer between friction, radiation transfer by the atmosphere and the body and thermal ablation; this gives us the possibility to examine the effects on the planetesimals and the atmosphere.}
   {We find that a significant amount of gravitational energy is indeed dissipated into the atmosphere, especially for larger planetary cores, which consequently cannot contribute to the atmospheric planetary luminosity. Furthermore, we examine that planetesimal infall events for cores,  $M_\mathrm{C} > 2$M$_{\oplus}$, which actually result in a local cooling of the atmosphere; this is totally in contradiction with the classical model.}
   {}


   \maketitle

\section{Introduction}
\label{sec_intro}
At early stages of planet formation, coagulation of dust and ice triggers solidification of bodies within the circumstellar gas disk that is present in the first few Myr. Even though the process behind the growth of such bodies is not well understood \citep{Morbidelli09, Johansen14}, if one accepts the core accretion scenario \citep{Perri74,Mizuno80}, protoplanets with several Earth masses, M$_{\oplus}$, have to exist at early disk stages. Once a planetary core reaches a sufficient mass, its gravitational potential dominates the local enthalpy of the surrounding disk gas and the gravitational accumulation of gas into an atmosphere begins.

In order to carry out time-dependent simulations for such atmospheres, it is a practical approach to assume hydrostatic and thermal equilibrium \citep{Stoekl15}. Hydrostatic equilibrium is fulfilled during most phases of atmospheric evolution, but thermal equilibrium is much more difficult to maintain. The absence of energy sources and sinks corresponds to a radially constant energy flow, and therefore a stationary atmospheric model requires a specification of the luminosity of a planet. The common approach in this instance is to associate the planetary luminosity with gravitational energy released by a flow of accreted planetesimals onto the surface of the protoplanet \citep{Hayashi79}. Despite the plausibility of such a luminosity source, this requires a quantification of the accretion rate and the distributions of planetesimal size and material strength.

The maximum lifetime of a circumstellar gas disk is typically about 10~Myr \citep{Williams11}, which implies a long-term mean mass accretion rate depending on the formation timescale $\mathcal{M}_{acc}$ of the system. Assuming $\mathcal{M}_{acc}$ to be the upper limit of protoplanetary mass accretion, the accretion rate for an individual protoplanet then only depends on the core mass M$_\mathrm{C}$.

To assess the effects of the accretion of solid bodies onto the planet, both the total accretion rate and the size distribution of the incoming bodies must be estimated. Whereas the mass accretion rate is constrained from the formation timescale $\mathcal{M}_{acc}$, a planetesimal size distribution is difficult to obtain. Planetesimals form in the disk by collisional coagulation as dust aggregates drift towards the mid-plane of the disk and form a thin, dense layer of larger bodies \citep{Weidenschilling00}. Collisional growth continues and produces planetesimals with sizes up to 10~km within a timescale of a few thousand years in the habitable zone. The simulations of \cite{Weidenschilling00} suggest a size distribution that is dominated by a relatively small number of 100~km bodies after about 1~Myr at a distance of 1~AU. Adopting their results, we finds an power-law particle distribution that has a maximum number density for submillimetre-sized objects and some rare 100~km-sized planetesimals. Knowledge about the size distribution of planetesimals is important because small objects experience more deceleration as they travel through the atmosphere and hence are not able to penetrate into deep atmospheric layers. The amount of deceleration depends on the amount of atmosphere captured by the protoplanet. As a consequence, small objects dissipate their gravitational energy already in the optically thinner parts of the atmosphere and only sufficiently massive planetesimals contribute significantly to the luminosity of the planet in the deep stationary atmosphere. However, large objects are less common and therefore contribute less to impact statistics.

Planetesimal material properties are an important factor for how an object travels through the atmosphere given that the object can in some cases fragment and eventually dissolve. High temperatures close to the protostar in the centre of the early disk dissociate matter into its elements, especially into H, C, N, O, Mg, Fe, Si, and S. As the disk gas is cooling down, those atoms recombine to form molecules that eventually form into dust grains and subsequently into planetesimals \citep{Gautier05}. Observational data, for example from the Infrared Space Observatory (ISO), show that interplanetary dust particles (IDP) mainly consist of silicates \citep{Rietmeijer04,Bouwman01,Keller02,Molster03} and hence an appropriate conjecture for the composition of planetesimals is a stony, silicate mineral.

However, if we take the effects of planetesimal size and strength into account, we might find that the assumption of impact driven, constant luminosity is invalid. In this study, we attempt to obtain further information about how much energy from impacting planetesimals is dissipated in the atmosphere and therefore cannot be released at the planetary surface. We show that in the case of heavy protoplanets, even cooling of the atmosphere can occur. Additionally, we demonstrate that mass accretion of planetesimals is a highly efficient process and hence might require further consideration in future models of protoplanetary primordial atmosphere simulations.

In Section~2, we describe our model for the transport of solid bodies through a primordial, planetary atmosphere.
In Section~3, we verify our model by comparing observational data and other verified models of actual meteorite infall events into the Earth's atmosphere with our model data.
In Section~4, we present the results for infalling planetesimals into primordial atmospheres of  protoplanets of various sizes.
Finally, in Section~5 we discuss the outcome of our simulations and give an overview about the scientific relevance of our results.

\section{Model description}
\label{sec_model_desc}
From the time-dependent models developed by \cite{Stoekl15}, we see that primordial planetary atmospheres quickly develop into dense, hot, extended structures. It is interesting to examine the process of objects falling through such dense atmospheres in more detail. Our main aim is to investigate the energy dissipation for a specific type of bolide and to explore how much of the initial mass is lost while travelling through the atmosphere.

Assessing the planetesimal size distribution in the disk is difficult, especially at different stages of atmospheric evolution during the embedded phase. Hence we have used time-dependent planetesimal simulations \citep{Weidenschilling00} and picked a particle diameter spectrum from \mbox{$1 \leq D_0 \leq 10^7$~cm} that is plausible at all reasonable disk stages. The spectrum is approximately a power law and is described later in the text. Furthermore, we assume this distribution remains constant during the accretion phase.

Following \cite{Perri74} and \cite{Mizuno80}, we assume that protoplanets with masses between $0.1$ and 5~M$_{\oplus}$ exist within the inner disk early enough in order to maintain an atmosphere up the the Hill radius. Moreover, as noted, planetesimal material properties are an important factor for interactions of planetesimals with the atmosphere. Hence,  a strength value $\sigma$ is required to describe
the fragmentation of bolides properly. For our study, we set $\sigma=10^8$~dyn~cm$^{-2}$, which represents an averaged value for a stony, silicate mineral \citep{Rietmeijer04,Bouwman01,Keller02,Molster03}.

\subsection{Atmospheric model}
\begin{figure}
        \centering
        \includegraphics[scale=0.44]{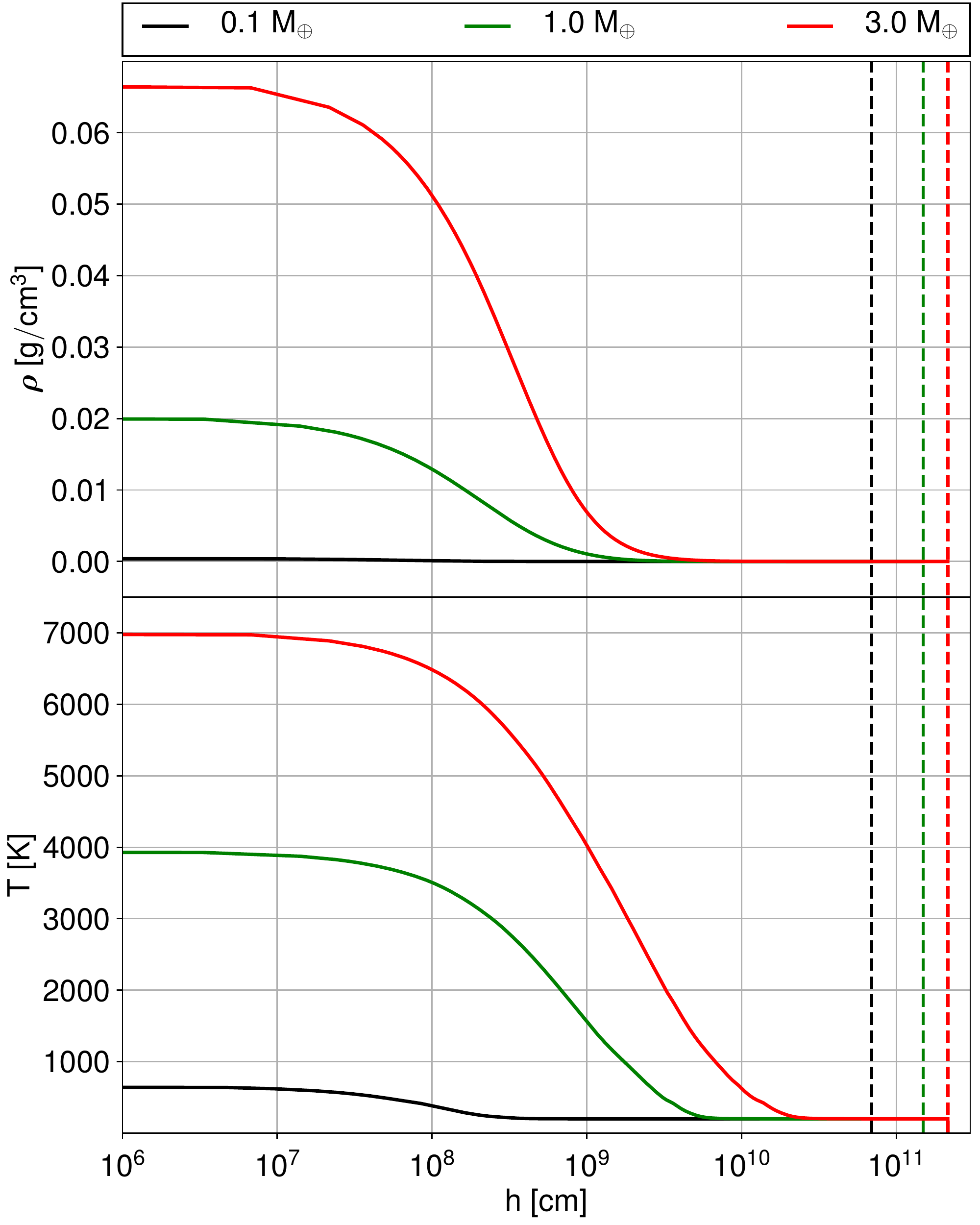}
        \caption{Density (upper panel) and temperature (lower panel) profiles of primordial atmospheres as a function of altitude, for core masses of 0.1~M$_\oplus$ (black), 1~M$_\oplus$ (green), and 3~M$_\oplus$ (red). The dashed lines represent the individual Hill radii of the cores. These data represent a snapshot at 1~Myr from the time-dependent atmospheric evolution simulations and a star with 1~M$_\odot$ \citep[see][]{Stoekl15}.}
        \label{fig_atmospheres} 
\end{figure}

When planetesimals cross the Hill radius $R_\mathrm{H}$, they start to interact with the primordial planetary atmosphere of the protoplanet. Therefore, our model adopts the density and temperature stratification of a corresponding atmospheric model. Generally, a protoplanet can accumulate gas from a circumstellar disk into a planetary envelope. As the structure and properties of nebula-embedded protoplanetary atmospheres are an inherently time-dependent problem, \cite{Stoekl15} developed a 1D spherically symmetric hydrodynamics code in order to simulate the accretion process of disk gas onto planetary cores and the subsequent evolution of embedded atmospheres. These authors considered core masses between 0.1 and 5~M$_{\oplus}$, situated in the habitable zone around a Sun-like star. It should be noted that the structure and composition of primordial atmospheres differ significantly from evolved planetary atmospheres that are no longer attached to the surrounding  disk medium. Such early atmospheres are much more massive and can reach extremely high surface temperatures (up to $10000K$). Fig.~\ref{fig_atmospheres} exhibits the density and temperature profiles for core masses with 0.1~M$_{\oplus}$, 1~M$_{\oplus}$, and 3~M$_{\oplus}$. Such atmospheres contribute several percent to the mass of the planet, whereas the Earth's atmosphere only adds about $10^{-6}\%$.

The maximum lifespan of a disk is estimated to be $\sim10$~Myr or between $1$ and 3~Myr for inner disks, respectively \citep{Williams11}. As we consider an embedded protoplanet, we utilized a snapshot at 1~Myr for various protoplanets ranging from \mbox{$0.1 \leq M_\mathrm{C} \leq 5$~M$_{\oplus}$} from the atmospheric models of \cite{Stoekl15}. Choosing snapshots at 1~Myr is, on the one hand, consistent with the existence of a disk (in the range of the common hypothesis of disk lifetime) and, on the other hand, such atmospheres appear to be in a dynamically stable state for a reasonable amount of time; there is no runaway accretion of disk gas over outer boundary and no significant turbulent flows in atmosphere (compare Fig.~\ref{fig_atmospheres}). The parameters that we use from the models of \cite{Stoekl15} are the atmospheric temperature, $T_\mathrm{atm}$, and the atmospheric density, $\rho_\mathrm{atm}$ as a function of the radius up to $R_\mathrm{H}$.

In the embedded phase, that is the phase during which the planet is embedded within the disk gas, the relative velocity of planetesimals with respect to the protoplanet is low because of the Keplerian motion of early disks. Hence we only consider head-on infall events with a small initial planetesimal velocity (relative velocity) $v_0$ at Hill Radius. In order to agree with the literature \citep{Kokubo12}, we set $v_0$ to the individual escape velocity at the surface for each bolide class. Different initial velocities affect the results but our simulations only show negligible variations in the outcome when setting the planetesimal infall speed to their surface escape velocity. We also note that for later phases, when the disk has vanished, the motion of planetesimals within $R_\mathrm{H}$ is dominated by the effects of the planet's gravity, and hence such infall scenarios do not apply correctly. Additionally, post-disk phase infall events require more detailed information about infall velocities and infall trajectories for appropriate simulation results.

\subsection{Time-dependent model}
\label{subsec_tdModel}
Consider a bolide with mass $m_\mathrm{ptm}$, at $R_\mathrm{H}$ with it's initial velocity $v_0$ for a protoplanetary mass $m_\mathrm{pl}$. As the body accelerates towards the protoplanet's surface, it collides with atmospheric gas, and the resulting atmospheric drag has two consequences for the bolide:
 First, the body heats up due to friction with the atmospheric gas, and second, the body breaks up if the ram pressure exceeds the bolide strength.
The exact fraction $f$ of heat transfer from atmospheric gas to the bolide is uncertain, but the upper limit of $f$ is given by $f=c_\mathrm{D}/2$, where $c_\mathrm{D}$ denotes the drag coefficient of the planetesimal \citep{Allen62}. \cite{Podolak88} argued that this value may in fact be considerably smaller ($f=c_\mathrm{D}/20$) and it certainly has to fulfil the condition $0<f_\mathrm{ene}<1$. By analysing observed meteorite falls (see Sec.~\ref{sec:model_verification}), we found that a value of $f_\mathrm{ene}=0.1$ best corresponds with the observational results; if $f_\mathrm{ene}>0.1$ the meteors were burned up too early, and if $f_\mathrm{ene}<0.1$, they did not show comparable mass loss when hitting the Earth's surface.

Friction heating not only causes the body to heat up but also results in a mass loss of surface material via sublimation and via melting and evaporation. This process is called thermal ablation and is discussed in detail by \cite{Podolak88}. Thermal ablation only affects the surfaces of infalling bodies because heat conductivity is slow compared to the ablation rate and thus leaves their core regions almost unaffected. For simplicity, and because the melting and evaporation temperatures of silicates are within the same order of magnitude, we do not distinguish between sublimation and melting followed by evaporation.

\begin{table*}
\centering
\caption{Bolide material constants used for all simulations.}
\label{tab_matconst}
\begin{tabular}{@{}lclll@{}}
\toprule\toprule
Parameter                               &   & Value                                       & Unit                 &  Info                             \\ \midrule
$T_\mathrm{m}$          & = & $1800$                 & [K]                      &  melting temperature \\
$c_\mathrm{D}$          & = & $1$                    &                  &  drag coefficient    \\
$\rho_\mathrm{ptm}$     & = & $3.2$                  & [g~cm$^{-3}$]    &  bolide density      \\
$H_\mathrm{f}$          & = & $3.041 \times 10^{10}$ & [erg~g$^{-1}$]   &  heat of fusion      \\
$\sigma_\mathrm{ptm}$   & = & $10^8$                 & [dyn~cm$^{-2}$]  &  bolide strength        \\ \bottomrule  
\end{tabular}
\end{table*}

Furthermore, we consider all bolides to be spherical and the material properties that we assume are listed in Table~\ref{tab_matconst}, where $H_\mathrm{f}$ denotes the amount of energy necessary to change a mass of 1~g from the solid to the liquid state, without changing its temperature.

Deriving a set of equations that fully describes the energetic state of the bolide is straight forward. The position of the body depending on the time spent travelling through the atmosphere is given by the velocity of the bolide
\begin{equation}
        \frac{d r}{d t} = v_\mathrm{ptm}
,\end{equation}
where $r$ represents the distance from the centre of the protoplanet.

In order to derive the equation of motion we consider two forces acting on the body. These are the gravitational force $F_\mathrm{G}$ and the resistance due to interaction with the atmosphere $F_\mathrm{D}$. 
Additionally, we assume the bolide velocity vector to point in the same direction as the gravitational force, which leads to
\begin{equation*}
        F_\mathrm{G} + F_\mathrm{a} - F_\mathrm{D} = 0,
\end{equation*}
where
\begin{align*}
        F_\mathrm{G} & = -G \frac{m_\mathrm{ptm} M}{r^2} = -GM \frac{\pi D_\mathrm{ptm}^3 \rho_\mathrm{ptm}}{6r^2} \\
        F_\mathrm{a} & = - m_\mathrm{ptm} a = - \frac{1}{6}\pi D_\mathrm{ptm}^3 \rho_\mathrm{ptm} \frac{d v_\mathrm{ptm}}{d t} \\
        F_\mathrm{D} & = \frac{1}{2}\rho_\mathrm{atm} v^2 c_\mathrm{D} A_\mathrm{proj} = \frac{1}{8}\pi D_\mathrm{ptm}^2 c_\mathrm{D} \rho_\mathrm{atm} v_\mathrm{ptm}^2,
\end{align*}
where $A_\mathrm{proj} = \frac{1}{4}\pi D_\mathrm{ptm}^2$ denotes the projected surface of the bolide and $P_\mathrm{ram} = \frac{1}{2}\rho_\mathrm{atm}v^2$ is the ram-pressure.
Combining these, we can obtain the equation of motion
\begin{equation}
        \frac{d v_\mathrm{ptm}}{d t} = -\frac{G M}{r^2} + \frac{3}{4} \frac{\rho_\mathrm{atm}}{\rho_\mathrm{ptm}} \frac{c_\mathrm{D}}{D_\mathrm{ptm}}v_\mathrm{ptm}^2.
\end{equation}

Concerning the energy balance of bolides and their exchange with the atmosphere, we consider the following processes:
\begin{itemize}
        \item friction heating
        \item radiation of bolide
        \item radiation of atmosphere
        \item thermal ablation,
\end{itemize}
whereas friction heating and the radiation of the bolide contribute to the heating of the atmosphere (energy input, $E_{in}$) and the radiation of the atmosphere itself plus thermal ablation are cooling and therefore take energy out ($E_{out}$) of the protoplanetary atmosphere.

\subsubsection*{Energy input into the atmosphere}

We consider two mechanisms for the heating of the atmosphere by the infalling body. These are radiation from the body and frictional heating. Since bolides radiate approximately as black bodies, the heating rate due to radiation is given by the Stefan-Boltzmann law\begin{equation}
        \label{eq_Etb}
        e_{T_\mathrm{ptm}} = \sigma_\mathrm{S} T_\mathrm{ptm}^4 \pi D_\mathrm{ptm}^2.
\end{equation} 
The frictional heating is given by the drag force multiplied by the speed of the body, i.e.
\begin{equation}
        \label{eq_Ed}
        e_\mathrm{D,atm} = F_\mathrm{D} v_\mathrm{ptm} ( 1 - f_\mathrm{ene} ) = \frac{1}{8} \pi D_\mathrm{ptm}^2 c_\mathrm{D} \rho_\mathrm{atm} v_\mathrm{ptm}^3 ( 1 - f_\mathrm{ene} ).
\end{equation}
The kinetic energy that the body loses from the drag force is given both to the atmosphere and to the body as thermal energy.
The factor $f_\mathrm{ene}$ is the fraction of this energy that is given to the body.

\subsubsection*{Planetesimal heating}
\label{subsubsec_Eout}
We consider two mechanisms for planetesimal heating. These are from the radiation of the atmosphere and the frictional drag. We estimate the radiative heating by assuming that the atmosphere radiates as a black body, meaning
\begin{equation}
        \label{eq_Eta}
        e_{T_\mathrm{atm}} = \sigma_\mathrm{S} T_\mathrm{atm}^4 \pi D_\mathrm{ptm}^2,
\end{equation}
The heating rate due to the drag force is 
\begin{equation}
        \label{eq_Ed2}
        e_\mathrm{ptm,D} = F_\mathrm{D} v_\mathrm{ptm} f_\mathrm{ene} = \frac{1}{8} \pi D_\mathrm{ptm}^2 c_\mathrm{D} \rho_\mathrm{atm} v_\mathrm{ptm}^3 f_\mathrm{ene}.
\end{equation}

We only take thermal ablation due to melting the body's surface regions into account. This gives the energy input rate that corresponds to the mass loss of the bolide as follows:
\begin{equation}
        \label{eq_Em}
        e_\mathrm{m} = -H_\mathrm{f} \frac{dm}{dt} = -\frac{1}{2}H_\mathrm{f} \rho_\mathrm{ptm}\pi D_\mathrm{ptm}^2 \frac{d D_\mathrm{ptm}}{d t}.
\end{equation}
Combining Eqn.~\ref{eq_Etb}, \ref{eq_Ed2}, \ref{eq_Eta}, and \ref{eq_Em} and considering the fact that energy input and energy output have to be in equilibrium, we can derive an equation that represents the diameter change due to mass loss of the bolide as follows:
\begin{equation*}
        \frac{d D_\mathrm{ptm}}{d t} = -\frac{1}{\rho_\mathrm{ptm} H_\mathrm{f}} \underbrace{\left[ \frac{1}{4} \rho_\mathrm{atm} c_\mathrm{D} f_\mathrm{ene} v_\mathrm{ptm}^3 + 2 \sigma_\mathrm{S} (T_\mathrm{atm}^4-T_\mathrm{ptm}^4)\right]}_{(\star)}.
\end{equation*}
As the surface of the body can only heat up until it reaches melting temperature before mass loss occurs, the surface temperature of the bolide has to be limited by its melting temperature $T_\mathrm{m}$. Therefore we set $(\star) = 0$ (boundary condition) and write for $T_\mathrm{ptm}$
\begin{equation*}
        T_\mathrm{ptm} = \min\left( \left[T_\mathrm{atm}^4 + \frac{1}{8 \sigma_\mathrm{S}}\rho_\mathrm{atm}c_\mathrm{D} f_\mathrm{ene} v_\mathrm{ptm}^3 \right]^{\frac{1}{4}}, T_\mathrm{m} \right).
\end{equation*}
Furthermore, we assume that the infalling material is composed of silicate, hence we set the melting temperature in our simulations to $T_\mathrm{m} = 1800$~K \citep{Podolak88}.

In addition to thermal ablation, the body can also break up if the ram pressure exceeds the internal bolide strength (strength criterion). We assume that break-up is a discrete event where the body separates into a certain specified number of fragments $n_\mathrm{sep}$ whenever the strength criterion is fulfilled. In our simulations, we set this fragmentation number to the lower limit of $n_\mathrm{sep} = 2$. After a break-up, the fragments do not act as individual bodies right away but only after a period of time in which the fragments drift apart (distance criterion). We consider that a body has broken apart when 
\begin{equation*}
\sigma_\mathrm{ptm} \le \frac{1}{2}\rho_\mathrm{atm} v_\mathrm{ptm}^2
\end{equation*}
and we consider the bodies to be separate bodies when  
\begin{equation*}
x_\mathrm{H} \ge f_\mathrm{dist} D_\mathrm{ptm,}
\end{equation*}
where $x_\mathrm{H}$ is the horizontal distance between the fragments and $f_\mathrm{dist}$ is an arbitrary factor that describes how far the fragments need to drift apart (in bolide diameters) until they are seen as individual objects. Obviously, an accurate description of how the fragments drift apart is not trivial \citep[see e.g.][]{Shuvalov14}. As our model does not need detailed information about the drift process, but only needs information about the drift velocity of the fragments, we followed \cite{Shuvalov14} who suggested that a good assumption for the horizontal drift velocity is
\begin{equation*}
        v_\mathrm{H} = |v_\mathrm{ptm}| \sqrt{\frac{\rho_\mathrm{atm}}{\rho_\mathrm{ptm}}}
\end{equation*}
and a distance parameter of $f_\mathrm{dist} = 2$ is a reasonable value to treat fragments as separate objects.

After complete break-up, we have $n_\mathrm{sep}$ bolides with diameters of $D_\mathrm{ptm} = D_\mathrm{ptm,0} / \sqrt[3]{n_\mathrm{sep}}$, where $D_\mathrm{ptm,0}$ denotes the bolide diameter before break-up. Furthermore, since the horizontal drift velocity is evaluated by the radial bolide velocity, it can be easily seen that the change of $v_\mathrm{H}$ is very small compared to the time it takes for the fragments to achieve the distance criterion. Thus we can assume that it is constant during the drift event, which implies that the diameter change can be assumed to be linear with separation distance $x_\mathrm{H}$. Hence the actual bolide diameter is
\begin{equation*}
        D_\mathrm{ptm}(t) = D_\mathrm{ptm,0} - \frac{1 - n_\mathrm{sep}^{-1/3}}{f_\mathrm{dist}} x_\mathrm{H},
\end{equation*}
with the corresponding time derivative
\begin{equation*}
        \frac{d D_\mathrm{ptm}}{d t} = - \frac{1 - n_\mathrm{sep}^{-1/3}}{f_\mathrm{dist}} v_\mathrm{H}.
\end{equation*}
Adding this finding to the equation of bolide mass loss we get
\begin{equation}
        \begin{split}
                \frac{d D_\mathrm{ptm}}{d t} = & -\frac{1}{\rho_\mathrm{ptm} H_\mathrm{f}} \, \max\left[ \frac{1}{4} \rho_\mathrm{atm} c_\mathrm{D} f_\mathrm{ene} v_\mathrm{ptm}^3 + 2 \sigma_\mathrm{S} (T_\mathrm{atm}^4-T_\mathrm{m}^4), 0\right] \\ 
                & - \frac{1-n_\mathrm{sep}^{-1/3}}{f_\mathrm{dist}}v_\mathrm{H}.
        \end{split}
\end{equation} 

\subsubsection*{Final set of equations}
Summarising all the results from this section the set of equations that have to be solved takes the following form
\begin{align}
        \frac{d r}{d t} = & v_\mathrm{ptm} , \\
        \frac{d v}{d t} = & -\frac{G M}{r^2} + \frac{3}{4} \frac{\rho_\mathrm{atm}}{\rho_\mathrm{ptm}} \frac{c_\mathrm{D}}{D_\mathrm{ptm}} v_\mathrm{ptm}^2 , \\
        \begin{split}
        \frac{d D_\mathrm{ptm}}{d t} = &
        -\frac{1}{\rho_\mathrm{ptm}H_\mathrm{f}} \, \max\left[ \frac{1}{4} \rho_\mathrm{atm} c_\mathrm{D} f_\mathrm{ene} v_\mathrm{ptm}^3 + 2 \sigma_\mathrm{S} (T_\mathrm{atm}^4-T_\mathrm{m}^4), 0\right] \\ 
        & - \frac{1-n_\mathrm{sep}^{-1/3}}{f_\mathrm{dist}} v_\mathrm{H}
        \end{split} , \\
        \frac{d x_\mathrm{H}}{d t} = &
        \begin{cases}
                v_\mathrm{H} & \text{at break-up,} \\
                0 & \text{otherwise,}
        \end{cases} 
\end{align}
which can be solved by numerical integration. We adopted a 4th order Runge-Kutta scheme with adaptive step-size control as described in \citet[page 554]{Press07}.

\subsection{Energy dissipation into a protoplanetary atmosphere}

Since we are interested in the total energy dissipation into the atmosphere, further considerations are necessary. The total energy dissipation is given by the sum of the energies dissipated by each of the bodies that travel through the atmosphere. Since the size and mass of an object determines how much energy it brings, it is necessary to consider the distribution of object sizes that enter the atmosphere. This distribution evolves over time, and has typically many millimetre-sized and a few kilometre-sized objects. In order to describe the particle size distribution, we followed \cite{Weidenschilling00}, whose simulations show that because of the growth of solids, the distribution is approximately following a power law. Using this result, we extract a number density per bolide diameter, per volume of surrounding disk material $N(D_\mathrm{ptm})$
\begin{equation}
\label{eq_distribution}
        \log(N(D_\mathrm{ptm})) = k\log(D_\mathrm{ptm}) + d,
\end{equation}
where $k$ is the slope and $d$ is the offset in the logarithmic size distribution.
Referring $N(D_\mathrm{ptm})$ on an initial bolide diameter $D_\mathrm{ptm,0}$ we get
\begin{equation*}
        N(D_\mathrm{ptm}) = N_\mathrm{ptm,0} \left( \frac{D_\mathrm{ptm}}{D_\mathrm{ptm,0}} \right)^k.
\end{equation*}
Adopting discrete mass bins (given by the bolide-diameter grid, where the index $i$ is used to indicate the quantity is for the $i$th diameter bin), the total mass per bolide class per volume of surrounding disk material is given by
\begin{equation*}
        \rho_\mathrm{ptm,i} = N_\mathrm{ptm,i} m_\mathrm{ptm,i} = \frac{1}{6}\frac{N_\mathrm{ptm,0}}{D_\mathrm{ptm,0}^k} \pi\rho_\mathrm{B} D_\mathrm{ptm,i}^{k+3}.
\end{equation*}

Mass accretion onto the protoplanet depends on the number of planetesimals within a given distance of the orbit of the planet. We estimate that the total mass of planetesimals that can effectively be accreted by the protoplanet is distributed in an annulus around the protostar. According to the work of \cite{Greenzweig1990}, the maximum radial extend of the accretion zone due to gravitational focussing is approximately 4~$R_\mathrm{H}$ (radial component) and the scale height $H_\mathrm{p}$ of the planetesimal disk can be estimated by $H_\mathrm{p} \sim 0.01 H$ \citep{Lambrechts2012}, where $H \sim 0.05 R_\mathrm{P}$ is the scale height of the gaseous disk  \citep{Armitage11}. This leads to a volume of the annulus of
\begin{equation*}
        V = 0.004 \pi R_\mathrm{P}^2 R_\mathrm{H},
\end{equation*}
where $R_\mathrm{H}$ is the Hill radius and $R_\mathrm{P}=1$~AU is the distance of the protoplanet to the protostar. Multiplying this by the individual densities gives the total masses within each diameter bin within the annulus. Summing the masses of all mass bins gives the total mass that is available for accretion onto the protoplanet as follows:
\begin{equation*}
        m_\mathrm{tot} = \sum_i M_\mathrm{ptm,i} = \sum_i \rho_\mathrm{ptm,i} V = \alpha \frac{N_\mathrm{ptm,0}}{D_\mathrm{ptm,0}^k} \sum_i D_\mathrm{ptm,i}^{k+3},
\end{equation*}
where $\alpha = 2\pi^2 \rho_\mathrm{B} R_\mathrm{P}^2 R_\mathrm{H} / 3000$.

Mass accretion does not happen instantaneously but instead happens over a certain period of time. Thus we can define a (mass-bin specific) frequency factor $\nu_\mathrm{i}$ by dividing the planet's mass accretion rate $\mathcal{M}_\mathrm{acc}$ by $m_\mathrm{tot}$ and multiplying with the weighted factor $\beta_\mathrm{i} = m_\mathrm{ptm,i} / m_\mathrm{tot}$, which gives
\begin{equation}
        \label{eq_freqFac}
        \nu_\mathrm{i} = \frac{\mathcal{M}_\mathrm{acc}}{m_\mathrm{tot}} \frac{m_\mathrm{ptm,i}}{m_\mathrm{tot}} = \frac{\mathcal{M}_\mathrm{acc}}{m_\mathrm{tot}} \beta_\mathrm{i},
\end{equation}
where $\nu_i$  describes the total infall rate per mass bin. The mass accretion rate for a specific planetary core can be determined by multiplying the core mass $M_\mathrm{C}$ (in Earth masses) with the mean mass accretion rate $\mathcal{M}_{\mathrm{acc},{\oplus}}=10^{-7} \text{M}_{\oplus} / yr$.

The total amount of energy dissipated in the protoplanetary atmosphere is given by the frictional heating plus the energy exchange rate between the bolide and the atmosphere (Section~\ref{subsec_tdModel}). As the object might break up while travelling through the atmosphere, the number of fragments must be taken into account. This value can be obtained by simply multiplying the number of fragments by the individual energy input rates per bolide mass bin,
\begin{align}
        e_\mathrm{fric,i}  & = N_\mathrm{ptm,i} (1 - f_\mathrm{ene})\frac{1}{8} \rho_\mathrm{atm} \pi c_\mathrm{D} v_\mathrm{ptm,i}^3 D_\mathrm{ptm,i}^2 \label{eq_efric} \\
        e_\mathrm{rad,i} & = N_\mathrm{ptm,i} \sigma_\mathrm{S} ( T_\mathrm{ptm,i}^4 - T_\mathrm{atm}^4 ) \pi D_\mathrm{ptm,i}^2 \label{eq_erad} \\
        e_\mathrm{diss,i} & = e_\mathrm{fric,i} + e_\mathrm{rad,i} \label{eq_etot},
\end{align}
where the number of fragments $N_\mathrm{ptm,i}$ is given by
\begin{equation*}
        N_\mathrm{ptm,i} = \frac{6 m_\mathrm{ptm,i}}{D_\mathrm{ptm,i}^3 \pi \rho_\mathrm{ptm}},
\end{equation*}
which can be obtained from the individual bolide mass using
\begin{equation}
        \label{eq_dm}
        \frac{d m_\mathrm{ptm,i}}{d t} = -\frac{\pi}{H_\mathrm{f}} D_\mathrm{ptm,i}^2 \,\max \left[ \frac{1}{8} \rho_\mathrm{atm} c_\mathrm{D} f_\mathrm{ene} v_\mathrm{ptm,i}^3 + \sigma_\mathrm{S} (T_\mathrm{atm}^4-T_\mathrm{m}^4), 0\right].
\end{equation}
It should be noted that to derive Eq.~ \ref{eq_dm}, melting in terms of thermal ablation is the only mass loss process we consider (no mass loss on break-up), so the above equation represents the conservation of mass for a particular bolide class.

Eqns.~\ref{eq_efric}, \ref{eq_erad}, and \ref{eq_etot} describe an energy input rate, hence, to compute the corresponding total amount of dissipated energy per bolide mass bin, these equations have to be integrated over time. For integration, we transform Eq.~\ref{eq_etot} to spatial coordinates according to $d / d r = v^{-1} d / d t$. Doing so, results in the total amount of dissipated energy by a single bolide mass bin being given by
\begin{multline}
        \label{eq_Eint}
        E_\mathrm{tot,i} = \int\limits_R N_\mathrm{ptm,i} \\
        \underbrace{\left[ \left(1-f_\mathrm{ene}\right)\frac{1}{8} \rho_\mathrm{atm} c_\mathrm{D} v_\mathrm{ptm,i}^2 D_\mathrm{ptm,i}^2 + \frac{\sigma_\mathrm{S}}{v_\mathrm{ptm,i}} \left( T_\mathrm{ptm}^4 - T_\mathrm{atm}^4\right)\right]}_{(\star \star)} dr,
\end{multline} 
where the term in braces is the total force per bolide diameter.
We integrate this equation numerically using a Simpson integration scheme.

In order to obtain the total amount of energy input into the atmosphere, the individual total energies (Eq.~\ref{eq_Eint}) have to be summed. Multiplying by Eq.~\ref{eq_freqFac}, leads to the energy input rate evaluated with the distribution function and mass accretion rate, given by
\begin{equation*}
        e_{\nu} = \nu E_\mathrm{tot} = \sum_i \nu_i E_{\mathrm{tot},i}.
\end{equation*}

We note that when we use too few mass bins, $e_{\nu}$ depends on the number of bins, which is undesirable. As we increase the number of bins, our calculated $e_{\nu}$ changes until $N_\mathrm{MB} > 700$, at which point we find that $e_{\nu}$ is independent of $N_\mathrm{MB}$. We therefore assume $N_\mathrm{MB}=701$, which give us an error less than $10\%$, which can be considered as an appropriate result in the limit of our model. Therefore, the computations in the next sections are performed with $N_\mathrm{MB}=701$.

\section{Model verification} \label{sec:model_verification}
Even though bolide dynamics in the Earth's atmosphere is not the primary objective of our model, observed meteorite falls are one valuable way to validate the reliability of our approach. Therefore a number of Earth related meteorite events with good observational records and/or simulation data were chosen and are listed in Table~\ref{tab_bolide_init}. Some of these data are based on estimates as neither remains of the objects were found nor any video material was available (e.g. Tunguska). Furthermore, as stated by some authors  \citep{Chyba93,Borovicka03}, even if video material is available, it is difficult to determine exact bolide parameters due to image saturation and/or fuzziness of the image.

Table~\ref{tab_bolide_init} shows the estimated initial parameters of the test meteorites, where $R_0$ is the initial height of the object and mainly corresponds to the starting point of observation, $M_0$ is the initial mass, $v_0$ is the initial velocity, $\rho$ is the density, and $\sigma$ is the strength of the meteorite. The other parameters are $h_\mathrm{AB}$, which represents the observed or modelled airburst height, $N_\mathrm{frag}$ the number of recovered fragments, and $M_\mathrm{rec}$ the total recovered mass of the bolide. These are the main indicators for our model verification.

\begin{table*}
\centering
\caption{Estimated properties of observed meteorite falls as presented by the literature from the given reference.}
\label{tab_bolide_init}
\resizebox{\textwidth}{!}{%
\begin{tabular}{@{}lccccccccc@{}}
\toprule\toprule
Bolide            & $R_0$ [km] & $M_0$ [kg]     & $v_0$ [km~s$^{-1}$] & $\rho$ [g~cm$^{-1}$] & $\sigma$ [dyn~cm$^{-2}$] & $h_\mathrm{AB}$ [km] & $N_\mathrm{frag}$ & $M_\mathrm{rec}$ [kg] & Referece            \\ \midrule
Tunguska          & $100$         & $5.6 \times 10^8$ & $15.0$          & $3.50$           & $1.0 \times 10^8$         & $5-10$           & $0$        & $0.00$             & \cite{Chyba93}     \\
Mor\'{a}vka     & $80$          & $1.5 \times 10^3$ & $22.5$          & $3.59$           & $5.0 \times 10^7$         & $29-37$          & $6$        & $1.40$             & \cite{Borovicka03} \\
Mor\'{a}vka MB  & $45.7$        & $1.1 \times 10^2$ & $21.9$          & $3.59$           & $5.0 \times 10^7$         & $29$             & $-$        & $-$                & \cite{Borovicka03} \\
Neuschwanstein    & $85$          & $3.0 \times 10^2$ & $20.9$          & $3.20$           & $1.1 \times 10^8$         & $21$             & $1$        & $1.75$             & \cite{Spurny03}    \\
Bunburra Rockhole & $62.8$        & $2.2 \times 10^1$ & $13.3$          & $2.70$           & $1.1 \times 10^6$         & $29.59$   & $3$        & $0.34$             & \cite{Spurny12}    \\
Bene\v{s}ov     & $r_\mathrm{H}$         & $3.0 \times 10^2$ & $21.0$          & $2.00$           & $1.6 \times 10^7$         & 35-40            & $0$        & $0.00$             & \cite{Artemieva01} \\ \bottomrule
\end{tabular}
}
\end{table*}

\subsection{Verification results}
There is no observational record (apart from seismic wave data and physical destruction of the landscape in the impact area) of the Tunguska event and therefore the initial parameters have to be chosen seperately. Because of this, we followed \cite{Chyba93} where different impact scenarios of the meteorite were studied. As our model only supports the perpendicular motion of bolides through the atmosphere, we only looked at those simulations with comparable set-ups. As there is no specific starting point mentioned for the Bene\v{s}ov bolide, we assume the initial height to be at Hill radius $R_\mathrm{H}$ in our computations.

\begin{figure}
        \centering
        \includegraphics[scale=0.44]{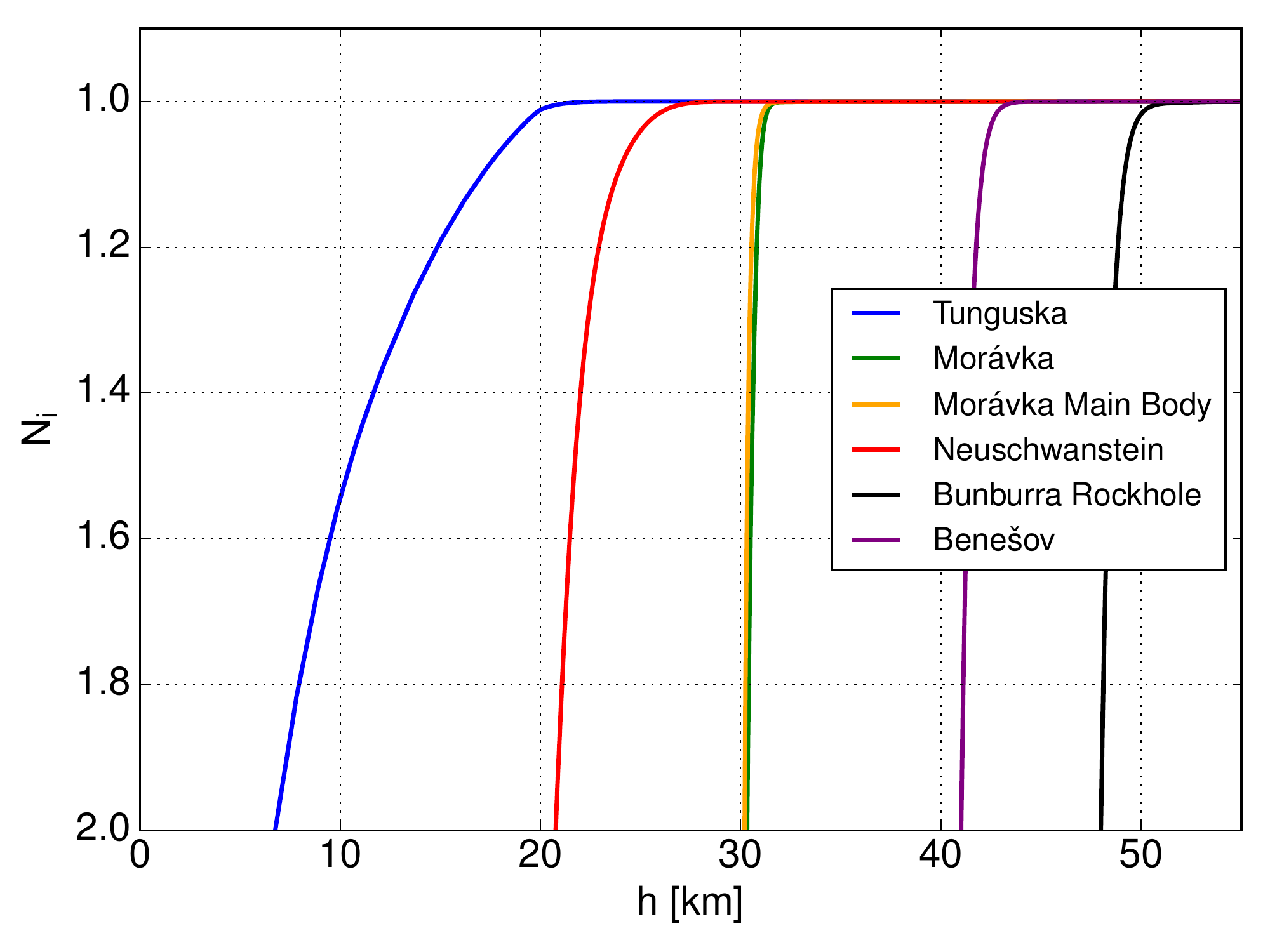}
        \caption{Airburst heights of the different meteorites; plot shows the number of fragments $N_i$ as a function of altitude. A number of fragments, $N_i=2$, indicates that both the strength criterion and distance criterion have been fulfilled for the first time. A number of fragments between 1 and 2 indicates that the strength criterion has been saisfied, but the distance criterion has not, and therefore we consider that the body has broken apart, but the fragments have not separated far enough to be considered separate bodies.}
        \label{fig_airburst}
\end{figure}

For the model of the Earth's atmosphere, we used the International Standard Atmosphere (ISA). Since ISA is only defined up to a height of 86~km and as there are some bolides with an initial height greater than that value, we isothermally extrapolated the atmospheric ISA values up to $R_\mathrm{H}$.  In our case, the choice of the atmospheric model (ISA, NRLMSISE, etc.) is not important because the density profiles in the lower atmosphere (up to 50~km) are comparable and densities above that altitude are low enough not to affect the meteorites aerodynamically (see Fig.~\ref{fig_airburst}).

Fig.~\ref{fig_airburst} shows airburst heights for the individual test candidates.\ The value $N_i=2$ indicates the first complete separation of two fragments (see Section~\ref{subsubsec_Eout}). One could assume that the actual fragmentation has already started as soon as the strength criterion is fulfilled (as soon as $N_i > 1$). But as the drift velocity of the fragments scale with the radial velocity of the bolide, the separation criterion is fulfilled soon after the strength criterion, compared to the distance the meteorite has travelled through the atmosphere. Hence we assume a fragmentation number of $2$ to be the indicator for the airburst event.

Comparing numerical results computed by our model with airburst heights stated by the various authors listed in Table~\ref{tab_bolide_init}, we can see that they are in good agreement, indicating that our model is a valid description of the actual physical processes. Almost all candidates show a break-up within the observed height or the values are close to the comparison models. Only the Mor\'{a}vka main body and the Bene\v{s}ov meteorite are slightly off. In the case of the Bene\v{s}ov event,  \cite{Artemieva01} referred to a reference particle with a slightly higher density of $\rho=4.2 \times 10^7$~dyn~cm$^{-2}$, which might affect the airburst height of our model. The computational results are listed in Table~\ref{tab_results} and the highest deviation from literature data for an airburst is registered for the Mor\'{a}vka main body with $3.1\%,$ which is still a good result considering a totally different model approach.

\begin{table}
\centering
\caption{Computed results for airburst heights for the different meteorite candidates. The model produces results in good agreement with the literature.}
\label{tab_results}
\begin{tabular}{@{}lcc@{}}
\toprule\toprule
Bolide                  & $h_\mathrm{AB,calc}$ [km]     &              \\ \midrule
Tunguska                & $6.80 \times 10^0$                    & $\checkmark$ \\
Mor\'{a}vka                     & $3.03 \times 10^1$                    & $\checkmark$ \\
Mor\'{a}vka MB          & $3.02 \times 10^1$                    & $\sim$       \\
Neuschwanstein          & $2.80 \times 10^1$                    & $\checkmark$ \\
Bunburra Rockhole       & $4.80 \times 10^1$                    & $\checkmark$ \\
Bene\v{s}ov             & $4.10 \times 10^1$                    & $\sim$       \\ \bottomrule
\end{tabular}
\end{table}

Comparing observational data other than airburst height must be treated with caution. This is because the number of found fragments $N_{frag}$ or the total recovered mass $M_{rec}$ are not reliable numbers as it is unlikely that all bolide fragments were found and recovered. In our case, it can be a good indicator for model verification in terms of whether a bolide impacts the surface or completely dissolves in the atmosphere, as we are not interested in detailed impact scenarios for the various meteorites.

\begin{figure}
        \centering
        \includegraphics[scale=0.44]{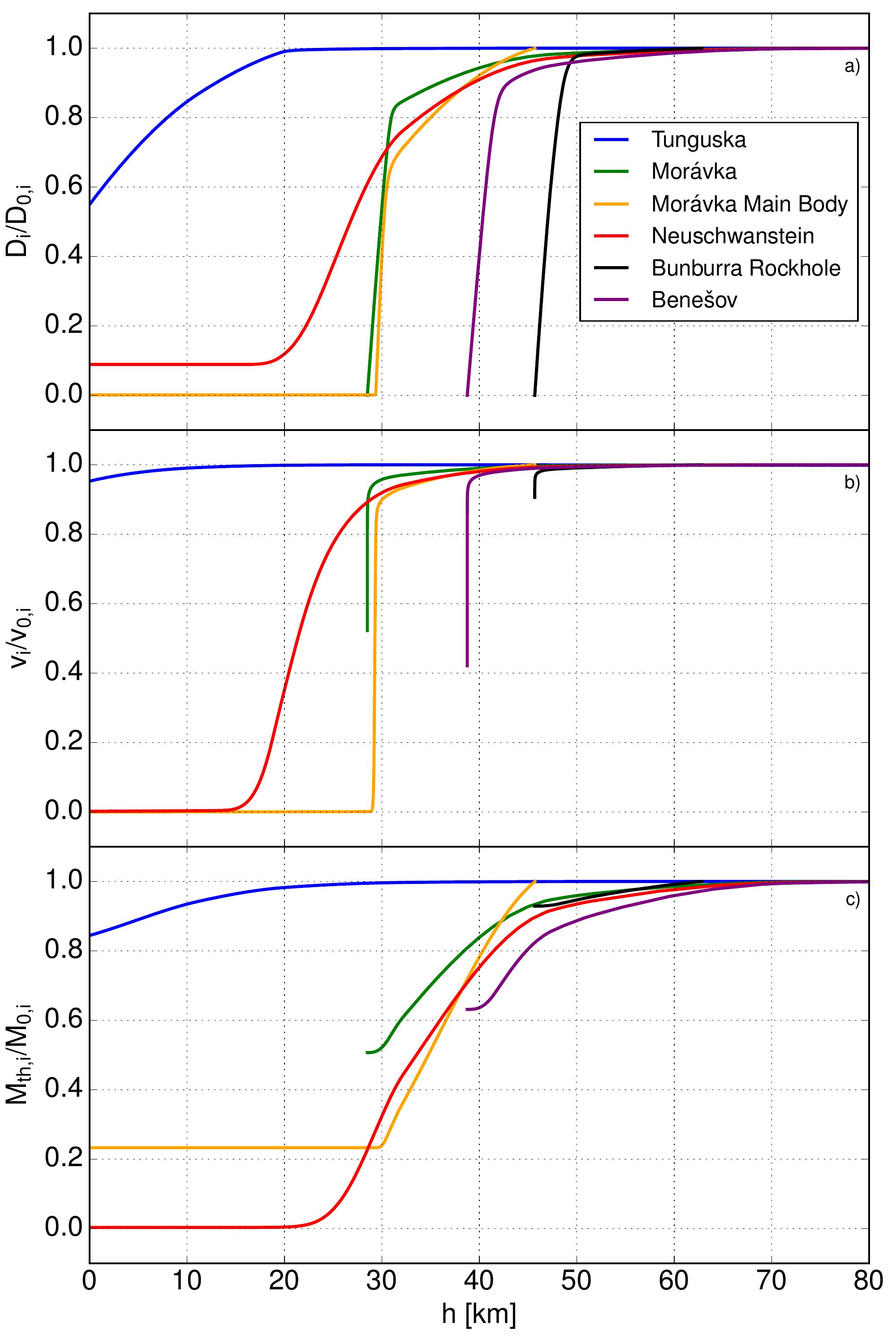}
        \caption{Combined plot for various meteorites. Panel $a)$ represents the normalised diameter reduction while travelling through the Earth's atmosphere, $b)$ the normalised velocity, and $c)$ the normalised mass loss as a function of altitude.}
        \label{fig_combined}
\end{figure}

Fig.~\ref{fig_combined} shows the bolide diameter, velocity, and mass while travelling through the atmosphere. All values are normalised to their initial values to obtain a more assessable plot. Steep slopes in the diameter plots represent runaway break-up, when multiple break-up events rapidly take place. This means that the break-up condition (strength criterion) is again fulfilled right after the fragments reach separation distance (distance criterion) and consequently trigger another fragmentation process. If the individual fragments cannot slow down quickly enough this process continues until the whole body completely dissolves in the atmosphere. In Fig.~\ref{fig_combined}, the Bene\v{s}ov event clearly shows these characteristics. In the simulation the bolide diameter drops quite rapidly down to submillimetre-sized dust particles that remain suspended in the atmosphere and eventually rain out after a certain amount of time. This corresponds to a fall in which no fragments of the meteorite were found.  

The Tunguska event clearly shows an impact of the bolide with a rather high velocity. Since the initial size of the bolide was large and the initial velocity rather low, there was no significant mass loss at the time of impact. Our model computes an impact energy (kinetic energy) of $4.85 \times 10^{23}$~erg that is in good agreement with the literature where the event is liberating between $4 \times 10^{23}$~erg \citep{Hunt60} and $4 \times 10^{25}$~erg \citep{Turco82}.

Our model suggests that in both of the Mor\'{a}vka events the remaining fragments would be around millimetre size. This does not agree with the observational data, but, as stated before, we can only simulate perpendicular motion, which does not compare with the literature as the trajectory is supposed to be around $20^\circ$ \citep{Borovicka03}. Furthermore, \citep{Borovicka03} describe the body as rather flat and this causes a higher deceleration in the upper atmosphere. In this case, our method does not allow for a detailed description of bolide behaviour after the first fragmentation event is not possible. Nevertheless, our model predicts a total recoverable mass of 760~kg, distributed over a large amount of fragments for the Mor\'{a}vka event. Simulation results for the Mor\'{a}vka main body \citep{Borovicka03}, predict a total mass of 17.1~kg at an altitude of 21.5~km. Comparing this result with our model data, we see that our model predicts a total mass of about 25~kg at the same altitude, which seems to be an acceptable result considering a completely different model approach.

Looking at the Neuschwanstein meteorite our model clearly predicts an impact that is in accordance with the literature. Our model suggests a total mass, possibly hitting the Earth's surface, of about 1.03~kg. This is a bit less than the mass of the recovered object with 1.75~kg \citep{Spurny03}. Again the authors describe the atmospheric trajectory of the bolide to be 49.5$^\circ$, which will have effects on the mass loss of the body.

The Bunburra Rockhole meteorite fall is a well observed event by the Australian Desert Fireball Network (DFN) \citep{Spurny12}. Our model suggests a catastrophic break-up for the meteorite down to submillimetre-sized particles. Considering the weak composition of the body, this again seems to be a reasonable result. Moreover, the heavy fragmentation triggers a large deceleration of the individual fragments, leading to a negligible amount of friction heating and thus no mass loss from thermal ablation. This seems to be in good agreement with the literature because only three rocks with a total mass of 339~g were found. Furthermore, the fragmentation events of the meteorite are described as ``explosive',' which substantiates very small fragments predicted by our model. Our model suggests a total mass, possibly hitting the surface, of $\sim$20~kg distributed over a large amount of submillimetre particles.

Summarising the above results, we emphasise that our model lacks a detailed reproduction of fragmentation after the initial break-up although it produces acceptable results with airburst heights for all cases. It does not allow us to analyse different atmospheric trajectories and does not take a shape factor into account, which has an impact on the aerodynamics of the object. It is not our aim in this paper to analyse detailed impact scenarios of bolides, but our focus lies on the energy dissipation in dense, protoplanetary atmospheres. With that in mind, our results are robust with respect to numerical and physical parameters and even deliberately choosing extreme values for planetesimal parameters only moderately affects the results.

\section{Results for protoplanetary atmospheres}
Our results for infalling planetesimals into primordial protoplanetary atmospheres reveal several interesting outcomes, some of which are seemingly counterintuitive. This underlines the demand of a reconsideration of the classical idea that all the kinetic energy of planetesimals is deposited on the surface of the planet. In this section, we apply our model to the case of bodies travelling through the thick protoatmospheres of planets embedded in the disk. We run a series of infall models for various body sizes and planetary masses.

\subsection{Effects on planetesimals}
\begin{figure*}
        \centering
        \includegraphics[scale=0.45]{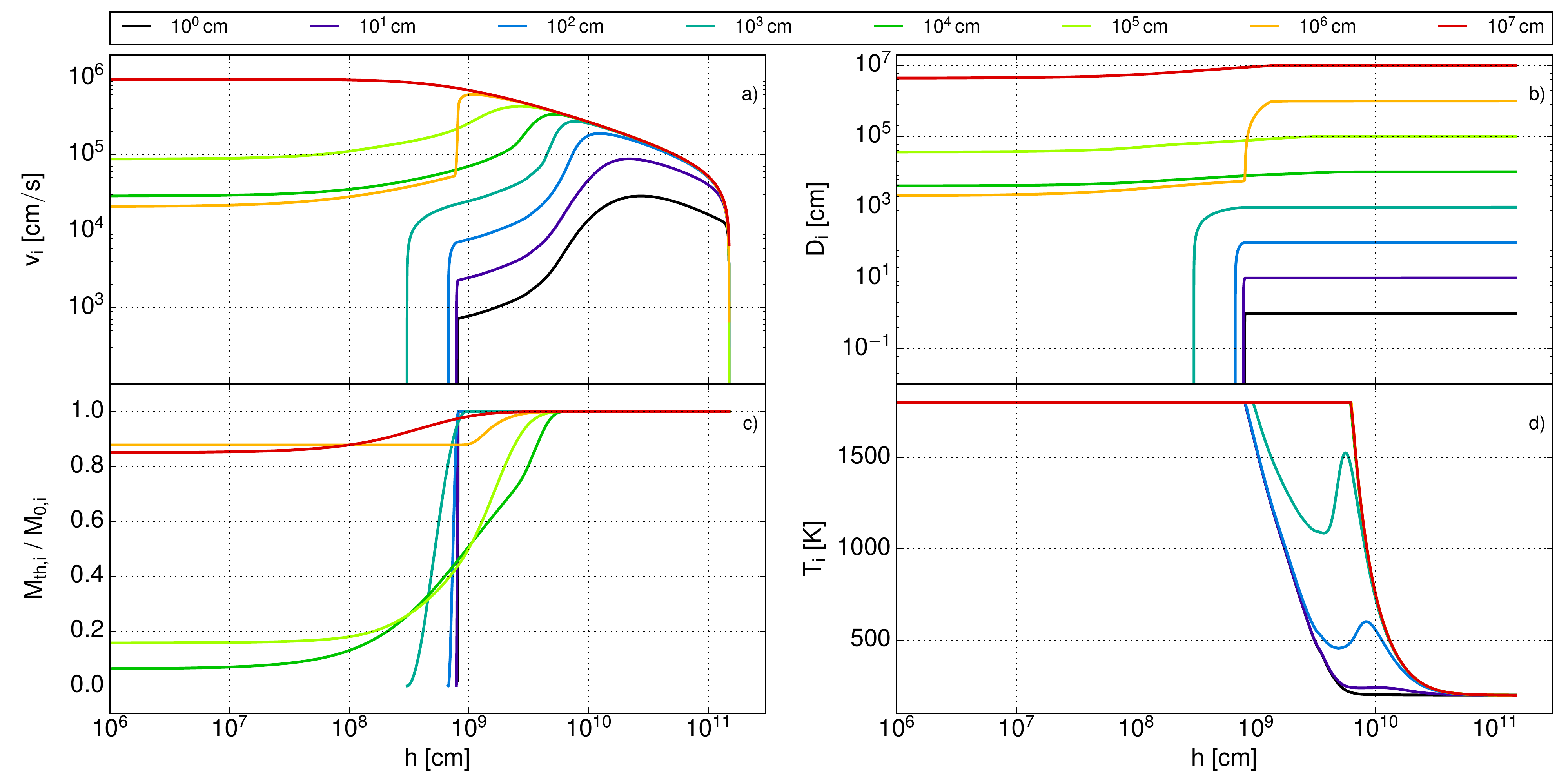}
        \caption{Figure comparing bolide velocity (panel a), diameter (panel b), mass loss (panel c), and temperature (panel d) as a function of altitude in the atmosphere for a planetary core mass of 1~M$_\oplus$ for various initial diameters ranging from $D_0=10^0$~cm to $D_0=10^7$~cm for one particular arbitrary chosen protoplanetary atmosphere.}
        \label{fig_multiplotM1}
\end{figure*}

Intuitively one might think that the more massive an infalling body, the more likely it will impact on the planetary surface and the higher its impact velocity, but inspecting Fig.~\ref{fig_multiplotM1}, the reality appears to be more complicated.

Fig.~\ref{fig_multiplotM1} shows a simulation for bolides with initial diameters ranging from $1 \leq D_0 \leq 10^7$~cm for a planetary core mass with 1~M$_\oplus$. All panels show eight logarithmically distributed test bodies starting from the lower limit (1~cm; black lines) to the upper limit (100~km; red lines). Even though our simulations were carried out all the way to the planetary surface, we cut all plots in Fig.~\ref{fig_multiplotM1} at an altitude of 10~km since below this altitude, there are no significant changes in the variables.

In Fig.~\ref{fig_multiplotM1}a, the 10~km test bolides eventually cross lines with bolides with lower masses and diameters, which implies that initially larger bodies do not always result in a higher impact velocity and vice versa. We see that small objects up to a size of 10~cm experience a deceleration already in the outer, rather tenuous atmosphere. Once the particle reaches the hot and dense layers (at around $h=10^9$~cm, cf. Fig.~\ref{fig_atmospheres}), gravitational acceleration and deceleration by friction, balance out and cause a flattening in the velocity curve. The relatively low speeds ($\sim 10$~m~s$^{-1}$) cause the particle to remain  in the atmosphere's high-temperature regions longer, which leads to a heating of the surface of the bolide and eventually, once the surface reached melting temperature (Fig.~\ref{fig_multiplotM1}d), to a complete dissolution by thermal ablation (Fig.~\ref{fig_multiplotM1}b and c).

Bodies with initial diameters from about 1~m to approximately 10~m are less affected by the outer atmosphere, thus reaching higher velocities. The higher velocity results in more friction heating leading to a higher surface temperature of the bolide. When reaching the inner, dense layers of the atmosphere, the planetesimal experiences a large deceleration. At this point, the temperature of the body is higher than the surrounding gas temperature and the decreasing efficiency of friction heating causes a radiation cooling of the bolide (negative gradients in Fig.~\ref{fig_multiplotM1}d). As the velocity decreases further and the atmospheric temperature rises, the  surface of planetesimal reaches melting temperature and, similar to smaller objects, the bolide disintegrates owing to thermal ablation.

Planetesimals with sizes around 100~m~$\leq D_0 \leq$~1~km are less affected by a deceleration in the outer atmosphere, but they experience more friction heating because of their larger effective surfaces. The velocity of such particles (in the denser atmosphere) is not enough to cause fragmentation but also high enough to cause significant friction heating and therefore extensive mass loss (Fig.~\ref{fig_multiplotM1}c).

Large planetesimals (10~km~$\leq D_0 \leq$~100~km) do not experiance much deceleration in the outer atmosphere, but again more friction heating owing to a larger cross section of the body. As one can see in Fig.~\ref{fig_multiplotM1}d, such bolide surfaces reach melting temperature rather early but owing to their high velocities, they do not remain in the hot regions long enough for a significant mass depletion (Fig.~\ref{fig_multiplotM1}c). Comparing the 10~km-sized with the 100~m-sized test bolide, we note that the 10~km object experiences a heavy deceleration to almost the same final velocity as the 100~m object. This behaviour is again attributable to the size of the object as the larger effective surface causes the planetesimal to experience a runaway break-up until the individual pieces are slow enough to dissatisfy the strength criterion (Section~\ref{sec_model_desc}).

\begin{figure}
        \centering
        \includegraphics[scale=0.44]{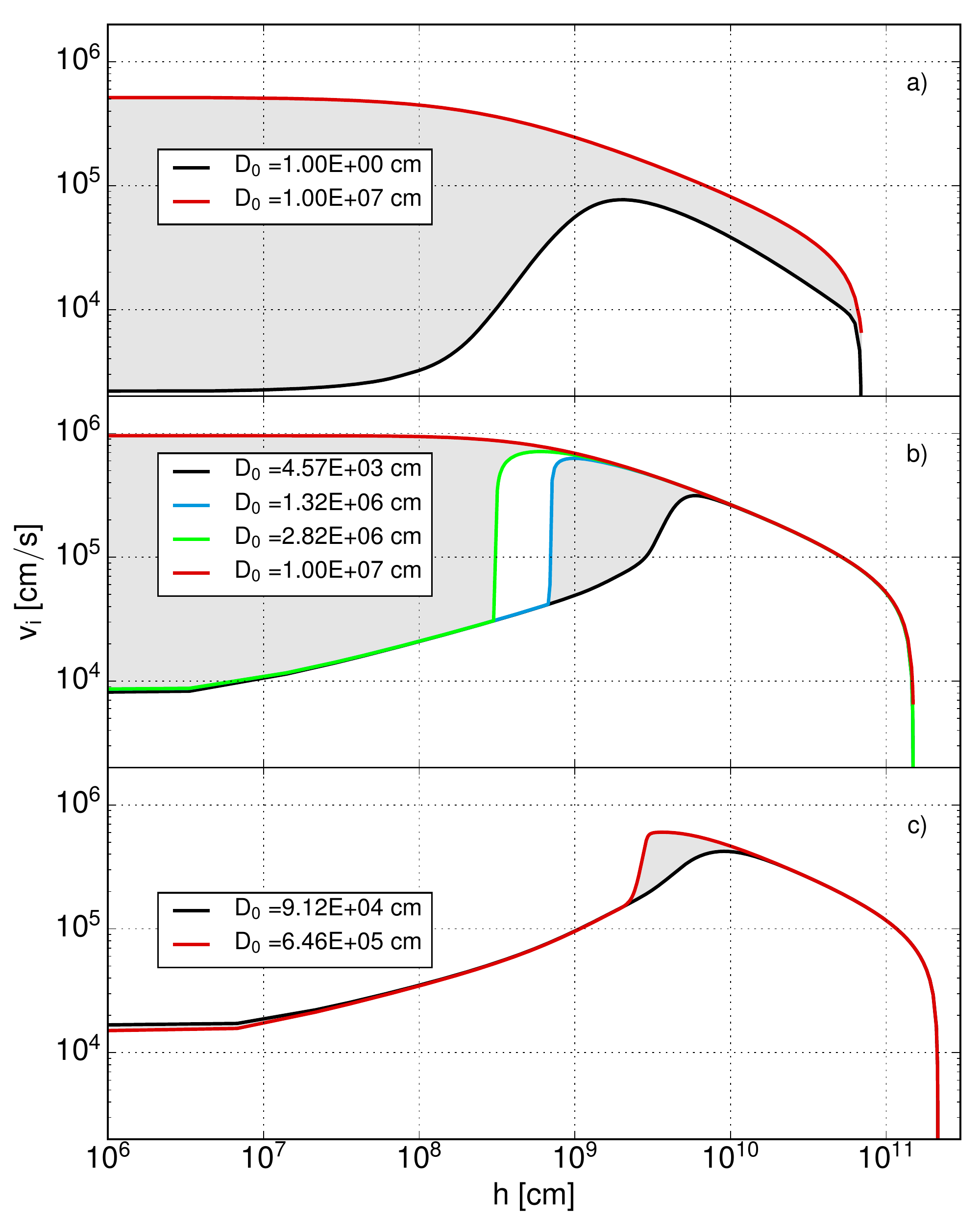}
        \caption{Diameter range of impacting planetesimals as a function of altitude for protoplanet core masses of 0.1~M$_\oplus$ (panel a), 1~M$_\oplus$ (panel b), and 3~M$_\oplus$ (panel c) as a function of altitude. Grey areas denote ranges between the lower and upper limit of impacting mass bins (solid lines). The plot shows bolide velocities $v_\mathrm{i}$ with respect to altitude $h$ with an impact velocity $v_\mathrm{h=0}>v_\mathrm{drift}$. All simulations are carried out with a grid of initial bolide diameters ranging from $1 \leq D_0 \leq 10^7$~cm.}
        \label{fig_impactors}
\end{figure}

If we investigate corresponding impact velocities (Fig.~\ref{fig_impactors}), we see again that bolide size is not necessarily an indicator for impacts. Some planetesimals are affected by a runaway break-up and might eventually dissolve in the atmosphere, whereas others are small enough to never exceed the strength criterion and, at the same time, are large enough so that mass loss due to thermal ablation is negligibly small.

Certainly, these processes depend very much on the structure and size of the protoplanetary atmosphere. Heavy planetary cores are surrounded by denser atmospheres that cause an earlier break-up for two equally sized bodies. Fig.~\ref{fig_impactors} shows impacting bodies for three protoplanets 0.1~M$_{\oplus}$, 1~M$_{\oplus}$, and 3~M$_{\oplus}$. Shaded areas represent the range of impactors at the protoplanetary surface with a speed higher than drift velocity $v_\mathrm{drift}$. The drift velocity is used as a criterion to stop the simulation when a particle reaches such low speeds that the time it remains in a quasi-isothermal atmospheric layer is comparable to the timescale of heat conduction. Protoplanetary atmospheres near the surface, reach temperatures much higher than melting temperatures of silicates \citep{Stoekl15}, which  cause a stony particle travelling at drift velocity to completely melt at one point (Fig.~\ref{fig_multiplotM1}c; bolide sizes up to some 10~m). Additionally, it appears to be reasonable that convective processes in such atmospheres are able to capture rather lightweight particles and trap them long enough to completely disintegrate the body. According to that, we set $v_\mathrm{drift}=1$~cm~s$^{-1}$, which seems to be an acceptable value taking the above statements into account. For the results represented in Fig.~\ref{fig_impactors} we ran simulations with the same, previously defined bolide grid but using a much finer grid of planetesimal sizes ($\Delta D = 10^{-2}$~cm).

A protoplanet with 0.1~M$_\oplus$ holds a much thinner atmosphere, at the same evolutionary phase, than a core with 1~M$_{\oplus}$. The smaller core mass accumulates less atmosphere during its evolution and has a smaller Hill radius. Therefore, not only is the initial gravitational energy for equally sized bolides less than for a larger protoplanet, but atmospheric drag is also weaker as a result of the thinner atmosphere. We see that in this case, the full range of planetesimals reaches the surface of the planet. Whereas large bodies (10~km to 100~km) are almost unaffected, small bodies experience a deceleration to about $v_\mathrm{ptm} \sim 10^3$~cm~s$^{-1}$ (Fig.~\ref{fig_impactors}a).

For a protoplanets with 1~M$_\oplus$, small bolides with sizes up to $D_0 \sim 45.7$~m, completely disperse in the dense, inner atmosphere (Fig.~\ref{fig_impactors}b). Larger planetesimals, up to the upper limit of the planetesimal grid, reach the surface either as a whole or as fragments with a smaller impact velocity (compare with Fig.~\ref{fig_multiplotM1}a).

Fig.~\ref{fig_impactors}c shows possible impact candidates for a protoplanet with 3~M$_\oplus$. As a result of a more extended, denser atmosphere, we are only left with a small gap of planetesimals ranging from $D_0=13.2$~km to $D_0=28.2$~km. As stated before, we count bodies as impactors if their surface velocity is higher than $v_\mathrm{drift}$. In the case of this scenario, all bodies (and their fragments) show relatively low impact speeds and hence cannot efficiently contribute to a core luminosity as the majority of their gravitational energy is already dissipated in the atmosphere.

\begin{figure}
        \centering
        \includegraphics[scale=0.43]{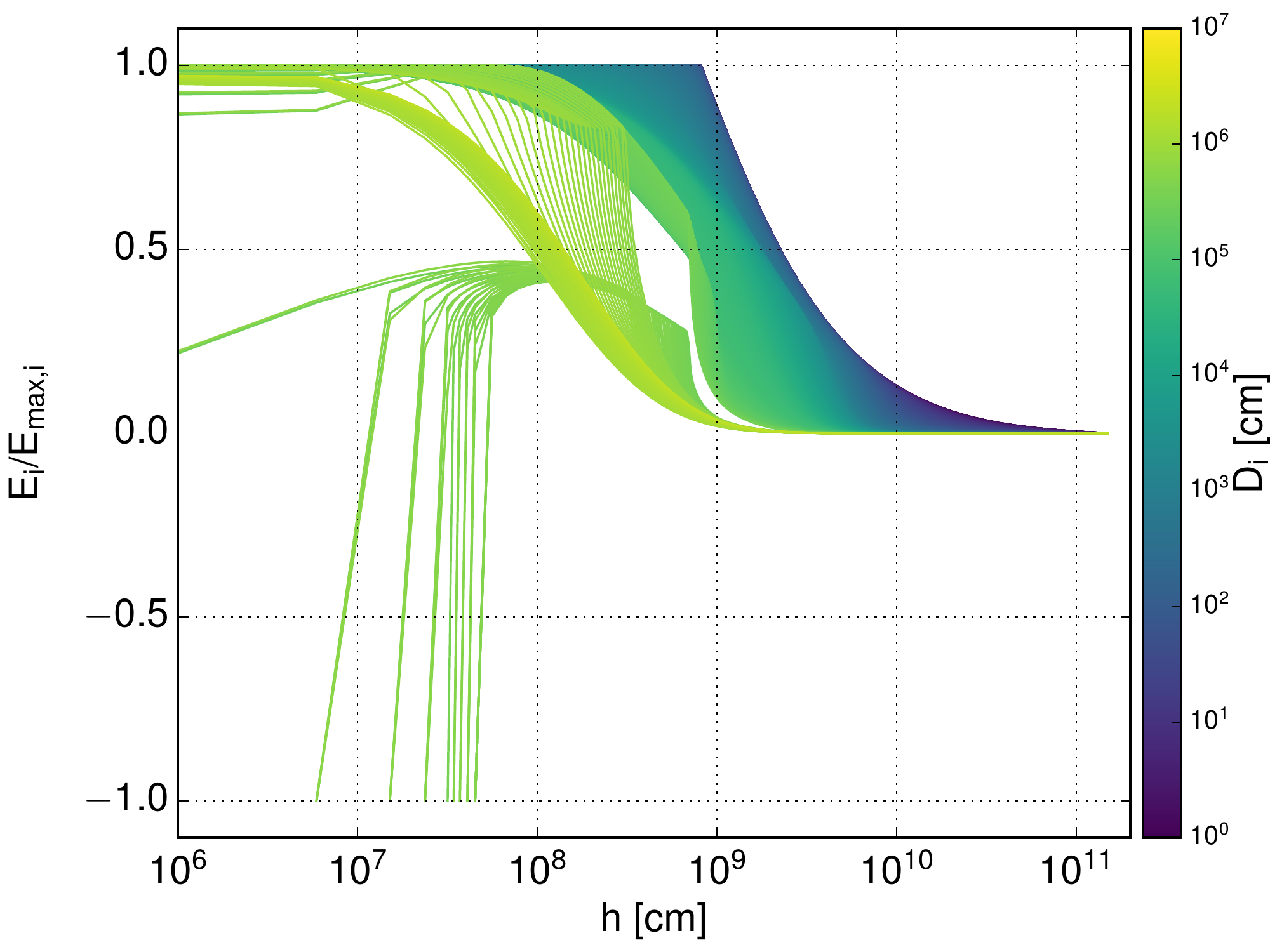}
        \caption{Energy dissipation per mass bin in units of $E_\mathrm{max}$ into the atmosphere of a protoplanet with 1~M$_\oplus$. The colours represent a bolide grid ranging from $D_0=1$~cm (purple) to $D_0=100$~km (yellow) with $N_\mathrm{B}=701$ mass bins. The total energy input is dominated by large bodies. Some bolide cool the atmosphere owing to phase transitions from solid to liquid.}
        \label{fig_energyDissipation}   
\end{figure}

\subsection{Effects on protoplanetary atmospheres}
Our main interest is the amount of energy that is dissipated while such objects travel through a protoplanetary atmosphere. Fig.~\ref{fig_energyDissipation} shows the relative energy dissipation for the same distribution of infalling bolides into a protoplanetary atmosphere with a planetary mass of 1~M$_\oplus$. The grid is represented by $N_\mathrm{B}=701$ mass bins with $1~\text{cm (purple)} \leq D_0 \leq 10^7~\text{cm (yellow)}$. We plot the relative total energy dissipation per mass bin, meaning that these results are not affected by the size distribution function. Small planetesimals (up to some 1~m; purple to blue lines) reach their maximum energy dissipation limit higher up in the atmosphere as they are more affected by atmospheric drag. Once they reach the inner dense layers, those objects start to disintegrate owing to thermal ablation. Since the transfer from the solid to liquid state of the bolide's surface layers requires a certain amount of energy ($H_\mathrm{f}$, Section~\ref{sec_model_desc}), the energy input into the atmosphere reduces by around $20\%$ for centimetre objects but there is no reduction for metre-sized objects. This finding represents the previously discussed, first gap (Fig.~\ref{fig_impactors}b) of non-impactors. Considering that such objects are small, the energy that is necessary for complete fusing is small and thus leading to a positive budget in energy input.

Larger bodies ($D_0 > 4.57 \times 10^3$~cm, Fig.~\ref{fig_impactors}b) appear with a negligible reduction of total energy input until we reach the lower limit of the second gap of non-impactors ($D_0 = 1.32 \times 10^6$~cm). This time, the total mass of the bolide is much higher and therefore much more energy for complete fusing is necessary. In the case of this specific bolide range (\mbox{$1.32 \times 10^6 \leq D_0 \leq 2.82 \times 10^6$~cm}), the total energy input is negative, resulting in an effective cooling of the atmosphere.

Once planetesimal size passes that second gap, the energy budget remains positive again. Even though those planetesimals break up into individual bodies, fragmentation occurs lower in the atmosphere and the individual fragments are fast enough to reach the surface without getting completely disintegrated.

\begin{figure}
        \centering
        \includegraphics[scale=0.44]{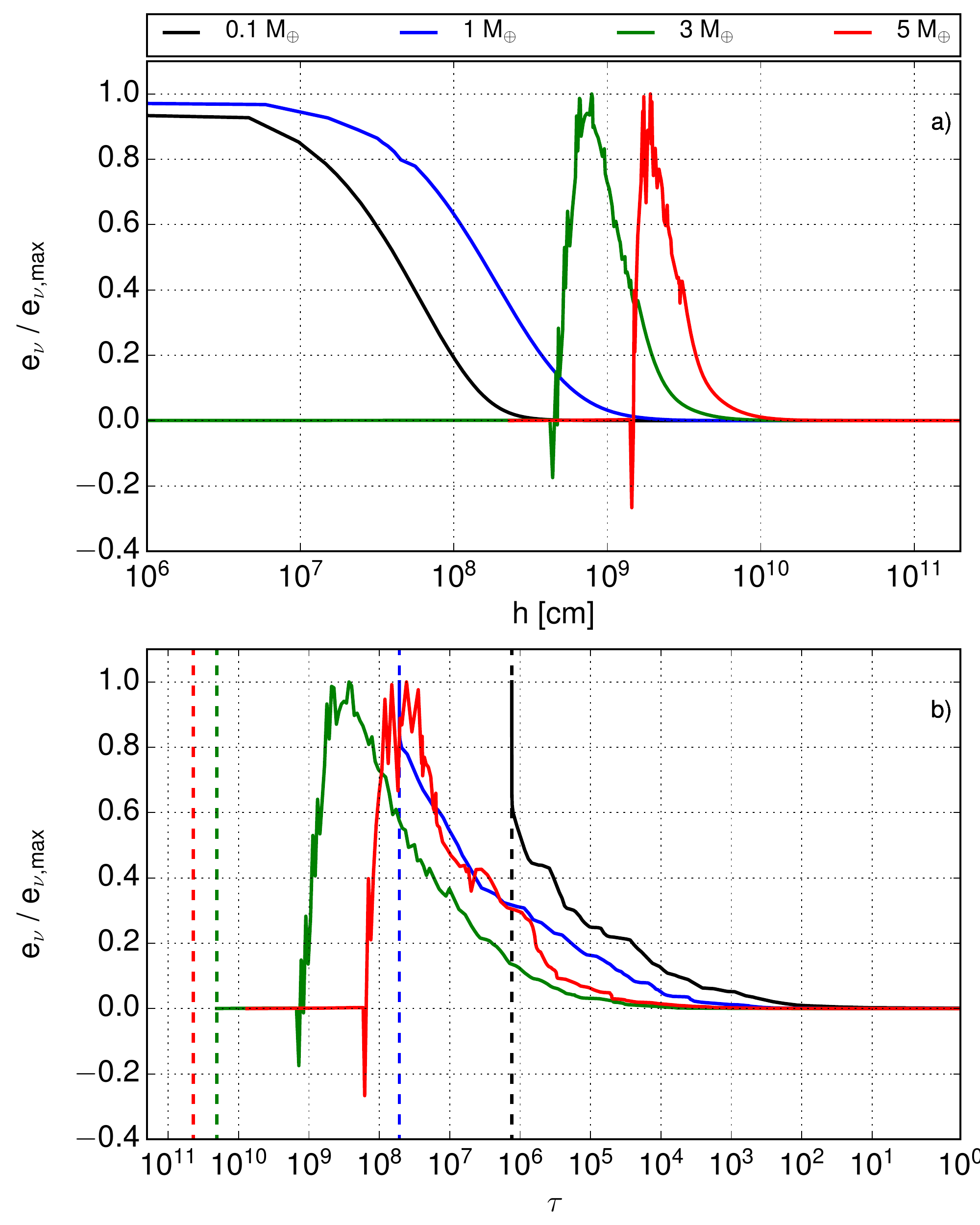}
        \caption{Panel a) shows the relative heating capacity $e_\mathrm{\nu}$ in values of $e_\mathrm{\nu,max}$ as a function of altitude for 4 different core masses M$_C=$~0.1~M$_\oplus$ (black), 1~M$_\oplus$ (blue), 3~M$_\oplus$ (green), and 5~M$_\oplus$ (red). Here $e_\mathrm{\nu}$ is the heating/cooling rate of the atmosphere evaluated with 701 mass bins, folded with the size distribution function $\nu$ and weighted with the planet's mass accretion rate $\mathcal{M}_\mathrm{acc}=10^{-7}$~M$_\mathrm{C}$ [M$_\oplus$/yr]. Whereas for smaller planets planetesimals cause a general heating of the atmosphere, for larger cores local cooling can occur (compare M$_\mathrm{C}=3$ and M$_\mathrm{C}=5$ M$_\oplus$).
        Panel b) shows $e_\nu$ in values of $e_\mathrm{\nu,tot}$ as a function of optical depth $\tau$. Again we compare 4 different core masses with M$_C=$~0.1~M$_\oplus$ (black), 1~M$_\oplus$ (blue), 3~M$_\oplus$ (green), and 5~M$_\oplus$ (red). The dashed lines represent the individual protoplanetary surfaces. Planetesimal infall results in a maximum energy input/output within the optical thick layers of the atmosphere.}
        \label{fig_energyInput} 
\end{figure}

Despite the fact that some of the large planetesimals locally extract energy from the atmosphere because of melting, the total energy input (folded with the size distribution function and evaluated with the relative mass accretion rate; Section~\ref{sec_model_desc}) for an $M_\mathrm{C} = 1$~M$_\oplus$ core is still positive (see Fig.~\ref{fig_energyInput}a, blue line) as the energy input is dominated by the largest bolides. Fig.~\ref{fig_energyInput}a shows the relative total energy input rate $e_{\nu,\mathrm{tot}}$ (see Section~\ref{sec_model_desc}) divided by the maximum energy input rate $e_{\nu,\mathrm{max}}$ as a function of altitude for four core masses: 0.1~M$_\oplus$ (black), 1~M$_\oplus$ (blue), 3~M$_\oplus$ (green), and 5~M$_\oplus$ (red). Whereas $e_{\nu,\mathrm{tot}}$ is overall positive for a planetary core with $M_\mathrm{C} = 1$~M$_\oplus$, this is not true for larger planetary cores. Heavier cores hold more atmosphere and consequently narrow the range of possible impactors (Fig.~\ref{fig_impactors}c) because of fragmentation. Hence, more planetesimal fragments remain in the atmosphere and heat up. Once the surface of the body reaches melting temperature, energy is required for phase transition, thus leading to a local cooling of the atmosphere. Our simulations show that protoplanetary atmospheres with cores larger than $M_\mathrm{C} > 3$~M$_\oplus$, capture $100\%$ of infalling planetesimals, which results in a significant cooling of lower atmospheric layers.

Finally, we discuss the results on how much energy is dissipated within the different layers of the atmosphere. Fig.~\ref{fig_energyInput}b shows the total heating rate $e_{\nu,\mathrm{tot}}$ in values of $e_{\nu,\mathrm{max}}$  with respect to the optical depth $\tau$, for core masses with $M_\mathrm{C}=0.1$~M$_\oplus$ (black), $M_\mathrm{C}=1$~M$_\oplus$ (blue), $M_\mathrm{C}=3$~M$_\oplus$ (green) and $M_\mathrm{C}=5$~M$_\oplus$ (red). We see that, regardless of core size, the maximum energy is dissipated in the optical thick part of the atmosphere, which implies that the interaction of planetesimals with protoplanetary atmospheres is an important factor of atmosperic evolution;  we note that we have cut the plot at $\tau = 10$ as below that
no significant changes in the values appear. Whereas with small planeteary cores most of the initial gravitational energy of the planetesimals is released as kinetic energy on impacts (compare Fig.~\ref{fig_impactors}), atmospheres around large cores cause heavy fragmentation even resulting in a local cooling of the atmosphere. As the optical thick layers of the atmosphere are fully convective, the atmosphere must react with expansion in case of heating and with contraction in case of cooling.

\section{Summary and discussion}
In this paper, we explore interactions of stony planetesimals with disk-embedded primordial protoplanetary atmospheres. Modelling a stationary atmosphere requires a specification of the  luminosity of the planet and a common approach is to associate luminosity with an impact energy of a stream of accreted planetesimals on the surface. Our main interest is to investigate how much gravitational energy is already depleted in the atmosphere and thus cannot contribute to a luminosity in the deep planetary atmosphere. Hence our simulations do not provide detailed impact statistics but give an overview about the total energy dissipation of accreting bodies of different initial sizes. We not only investigate the individual behaviour of such infalling objects but also the total energy dissipation as a result of a designated planetesimal distribution in the surrounding disk.

In Section~\ref{sec:model_verification}, we point out that our model lacks a detailed description of the fragmentation process of observed asteroids interacting with the Earth's atmosphere, although it produces acceptable results with airburst heights for all tested cases. Our model does not take non-radial atmospheric trajectories nor a shape factor into account, both of which can influence the aerodynamics of the object. Keeping this in mind, the results in Section~\ref{sec:model_verification} are robust with respect to numerical and physical limitations and even investigating the extreme of the possible planetesimal parameters only moderately changes the results. 

Protoplanets of different sizes accumulate different amounts of atmospheric gas during the disk phase. Primordial atmospheres are dense structures compare to secondary atmospheres (e.g. Earth's present atmosphere) and extend up to the Hill radius. Time-dependent simulations show us that the structure of such a primordial atmosphere consists of a thin, cold outer region and a dense, hot inner layer around the protoplanet. If we consider a constant planetary luminosity that is generated by undisturbed infalling planetesimals, it is important to consider how much of the gravitational energy of the planetesimals is actually lost by friction heating as a consequence of interactions with the atmosphere and subsequently cannot contribute to core luminosity. Our results clearly show that, depending on the core mass, a significant amount of energy is released within the atmosphere. This can be as much as 100\% for test bodies with diameters ranging from $1 \leq D_0 \leq 10^7$~cm, for a core mass with $M_\mathrm{C} > 3$~M$_\oplus$, with a primordial atmosphere at 1~Myr (as discussed in Section~\ref{sec_intro}). One could comment that for an Earth-sized core, the maximum energy dissipation is reached at only around $h=50$~km (see Fig.~\ref{fig_energyInput}; blue line) but this certainly does not apply for larger protoplanets. Furthermore, even though energy dissipation is low within the upper atmosphere, the dynamical effects are complicated and need further investigation.

An interesting result is that parts of the atmosphere of large protoplanets can also be cooled,  rather than heated, by planetesimal infall events. This clearly requires further analysis in future time-dependent models. Furthermore, partial cooling of the atmosphere could lead to contraction, leading to an accelerated inflow of surrounding disk gas and consequently to an accelerated mass increase. This scenario might be a contributing factor to why large gaseous planets already exist at early stages of disk formation. In order to get a better understanding of the dynamical effects in protoplanetary atmospheres caused by infalling planetesimals, implementing our model in the time-dependent simulations of \cite{Stoekl15} is necessary to investigate this hypothesis. Such an approach will give us the opportunity to study the time-dependent evolution of atmospheres including the back-reaction of planetesimals on the gas.

We want to emphasise that the atmospheric evolution models do not take planetary core growth into account \citep{Stoekl15}. Our model gives an overview about how much mass during various infall scenarios contributes to the atmosphere and how much reaches the surface. Additionally, we see that mass accretion for small bodies (up to $D_0=1$~m for $M_\mathrm{C}=1$~M$_\oplus$) is a very efficient process (Fig.~\ref{fig_multiplotM1}a) as those planetesimals experience drag already in the upper atmosphere and consequently slow down. This suggests that small-sized planetesimals remain trapped within the atmosphere regardless of infall speeds and infall trajectories. Considering the fact that the number density of such particles in the disk is significantly high, this mass gain should not be neglected. Additionally, we see a huge mass loss even for larger bolides (up to $D_0 = 10^5$~cm) while travelling through the atmosphere for an Earth-sized protoplanet (Fig.~\ref{fig_multiplotM1}c). This mass is indeed captured within the atmosphere and as we examine silicates, this mass cannot leave the atmosphere under normal circumstances. Hence those captured silicates eventually rains out onto the core once the atmosphere is cooling down. Looking into such scenarios gets even more important when studying larger protoplanets with $M_\mathrm{C}>1$~M$_\oplus$. 

Recent core growth simulations of giant gaseous planets \citep{Helled2008} suggest that planetesimal infall can drastically increase the results of the final mass of the planet due to adding a siginificant amount of refractory material. As our model not only considers the infall rate of plantesimals but can determine at which layer the planetesimal mass is depleted, our findings can be used to improve such core growth models.

Since core growth is an important question for planetary formation, including our model in time-dependent atmospheric evolution models will eventually result in a more accurate description of primordial protoplanetary atmosphere evolution. Such a consistent time-dependent model is not only relevant during the disk phase but also afterwards as long as the planet retains a primordial atmosphere.

In this context we also see similarities with the work of \cite{Movshovitz2010}, which focusses on the growth rate of giant planets due to coagulation of dust grains within the atmosphere. These authors consider infall rates due to planetesimal accretion, which is the source term for their coagulatiuon model. Nevertheless, they have no information about how much planetesimal mass is depleted in the various layers of the atmosphere. Our model predicts mass loss as a function of altitude, which could indeed enhance such coagulation models and give a better understanding of giant planet formation.

\begin{acknowledgements}
The authors acknowledge support by the FWF NFN project S116 ``Pathways to Habitability: From Disks to Active Stars, Planets and Life'', and the related subprojects S11601-N16 ``Hydrodynamics and Radiation in Young Star Disk Systems'' and S11604-N16 ``Radiation \& Wind Evolution from T Tauri Phase to ZAMS and Beyond''. This publication is supported by the Austrian Science Fund (FWF).
Furthermore we want to thank the referee for clarifying the presentation of some of our results.
\end{acknowledgements}

\bibliographystyle{aa}    
\bibliography{final}

\begin{thebibliography}{30}
\expandafter\ifx\csname natexlab\endcsname\relax\def\natexlab#1{#1}\fi

\bibitem[{{Allen}(1962)}]{Allen62}
{Allen}, H.~J. 1962, NASA Special Publication, 24, 1

\bibitem[{{Armitage}(2011)}]{Armitage11}
{Armitage}, P.~J. 2011, \araa, 49, 195

\bibitem[{{Artemieva} \& {Shuvalov}(2001)}]{Artemieva01}
{Artemieva}, N.~A. \& {Shuvalov}, V.~V. 2001, \jgr, 106, 3297

\bibitem[{{Borovicka} \& {Kalenda}(2003)}]{Borovicka03}
{Borovicka}, J. \& {Kalenda}, P. 2003, Meteoritics and Planetary Science, 38,
  1023

\bibitem[{{Bouwman} {et~al.}(2001){Bouwman}, {Meeus}, {de Koter}, {Hony},
  {Dominik}, \& {Waters}}]{Bouwman01}
{Bouwman}, J., {Meeus}, G., {de Koter}, A., {et~al.} 2001, \aap, 375, 950

\bibitem[{{Chyba} {et~al.}(1993){Chyba}, {Thomas}, \& {Zahnle}}]{Chyba93}
{Chyba}, C.~F., {Thomas}, P.~J., \& {Zahnle}, K.~J. 1993, \nat, 361, 40

\bibitem[{{Gautier} \& {Hersant}(2005)}]{Gautier05}
{Gautier}, D. \& {Hersant}, F. 2005, \ssr, 116, 25

\bibitem[{Greenzweig \& Lissauer(1990)}]{Greenzweig1990}
Greenzweig, Y. \& Lissauer, J.~J. 1990, Icarus, 87, 40

\bibitem[{{Hayashi} {et~al.}(1979){Hayashi}, {Nakazawa}, \&
  {Mizuno}}]{Hayashi79}
{Hayashi}, C., {Nakazawa}, K., \& {Mizuno}, H. 1979, Earth and Planetary
  Science Letters, 43, 22

\bibitem[{Helled \& Schubert(2008)}]{Helled2008}
Helled, R. \& Schubert, G. 2008, Icarus, 198, 156

\bibitem[{{Hunt} {et~al.}(1960){Hunt}, {Palmer}, \& {Penney}}]{Hunt60}
{Hunt}, J.~N., {Palmer}, R., \& {Penney}, W. 1960, Philosophical Transactions
  of the Royal Society of London Series A, 252, 275

\bibitem[{{Johansen} {et~al.}(2014){Johansen}, {Blum}, {Tanaka}, {Ormel},
  {Bizzarro}, \& {Rickman}}]{Johansen14}
{Johansen}, A., {Blum}, J., {Tanaka}, H., {et~al.} 2014, Protostars and Planets
  VI, 547

\bibitem[{{Keller} {et~al.}(2002){Keller}, {Hony}, {Bradley}, {Molster},
  {Waters}, {Bouwman}, {de Koter}, {Brownlee}, {Flynn}, {Henning}, \&
  {Mutschke}}]{Keller02}
{Keller}, L.~P., {Hony}, S., {Bradley}, J.~P., {et~al.} 2002, \nat, 417, 148

\bibitem[{{Kokubo} \& {Ida}(2012)}]{Kokubo12}
{Kokubo}, E. \& {Ida}, S. 2012, Progress of Theoretical and Experimental
  Physics, 2012, 01A308

\bibitem[{Lambrechts \& Johansen(2012)}]{Lambrechts2012}
Lambrechts, M. \& Johansen, A. 2012, Astronomy {\&} Astrophysics, 544, A32

\bibitem[{{Mizuno}(1980)}]{Mizuno80}
{Mizuno}, H. 1980, Progress of Theoretical Physics, 64, 544

\bibitem[{{Molster} \& {Waters}(2003)}]{Molster03}
{Molster}, F.~J. \& {Waters}, L.~B.~F.~M. 2003, in Lecture Notes in Physics,
  Berlin Springer Verlag, Vol. 609, Astromineralogy, ed. T.~K. {Henning},
  121--170

\bibitem[{{Morbidelli} {et~al.}(2009){Morbidelli}, {Bottke}, {Nesvorn{\'y}}, \&
  {Levison}}]{Morbidelli09}
{Morbidelli}, A., {Bottke}, W.~F., {Nesvorn{\'y}}, D., \& {Levison}, H.~F.
  2009, \icarus, 204, 558

\bibitem[{Movshovitz {et~al.}(2010)Movshovitz, Bodenheimer, Podolak, \&
  Lissauer}]{Movshovitz2010}
Movshovitz, N., Bodenheimer, P., Podolak, M., \& Lissauer, J.~J. 2010, Icarus,
  209, 616

\bibitem[{{Perri} \& {Cameron}(1974)}]{Perri74}
{Perri}, F. \& {Cameron}, A.~G.~W. 1974, \icarus, 22, 416

\bibitem[{{Podolak} {et~al.}(1988){Podolak}, {Pollack}, \&
  {Reynolds}}]{Podolak88}
{Podolak}, M., {Pollack}, J.~B., \& {Reynolds}, R.~T. 1988, \icarus, 73, 163

\bibitem[{Press {et~al.}(2007)Press, Teukolsky, Vetterling, \&
  Flannery}]{Press07}
Press, W.~H., Teukolsky, S.~A., Vetterling, W.~T., \& Flannery, B.~P. 2007,
  Numerical Recipes 3rd Edition: The Art of Scientific Computing, 3rd edn. (New
  York, NY, USA: Cambridge University Press)

\bibitem[{{Rietmeijer}(2004)}]{Rietmeijer04}
{Rietmeijer}, F.~J.~M. 2004, Earth Moon and Planets, 95, 321

\bibitem[{{Shuvalov} {et~al.}(2014){Shuvalov}, {K{\"u}hrt}, {de Niem}, \&
  {W{\"u}nnemann}}]{Shuvalov14}
{Shuvalov}, V., {K{\"u}hrt}, E., {de Niem}, D., \& {W{\"u}nnemann}, K. 2014,
  \planss, 98, 120

\bibitem[{{Spurn{\'y}} {et~al.}(2012){Spurn{\'y}}, {Bland}, {Shrben{\'y}},
  {Borovi{\v c}ka}, {Ceplecha}, {Singelton}, {Bevan}, {Vaughan}, {Towner},
  {McClafferty}, {Toumi}, \& {Deacon}}]{Spurny12}
{Spurn{\'y}}, P., {Bland}, P.~A., {Shrben{\'y}}, L., {et~al.} 2012, Meteoritics
  and Planetary Science, 47, 163

\bibitem[{{Spurn{\'y}} {et~al.}(2003){Spurn{\'y}}, {Oberst}, \&
  {Heinlein}}]{Spurny03}
{Spurn{\'y}}, P., {Oberst}, J., \& {Heinlein}, D. 2003, \nat, 423, 151

\bibitem[{{St{\"o}kl} {et~al.}(2016){St{\"o}kl}, {Dorfi}, {Johnstone}, \&
  {Lammer}}]{Stoekl15}
{St{\"o}kl}, A., {Dorfi}, E.~A., {Johnstone}, C.~P., \& {Lammer}, H. 2016,
  \apj, 825, 86

\bibitem[{{Turco} {et~al.}(1982){Turco}, {Toon}, {Park}, {Whitten}, {Pollack},
  \& {Noerdlinger}}]{Turco82}
{Turco}, R.~P., {Toon}, O.~B., {Park}, C., {et~al.} 1982, \icarus, 50, 1

\bibitem[{{Weidenschilling}(2000)}]{Weidenschilling00}
{Weidenschilling}, S.~J. 2000, \ssr, 92, 295

\bibitem[{{Williams} \& {Cieza}(2011)}]{Williams11}
{Williams}, J.~P. \& {Cieza}, L.~A. 2011, \araa, 49, 67

\end{thebibliography}

\end{document}